\documentclass[mathleft
]{an}
\usepackage{graphicx}
\usepackage{times}
\begin{document}


\title{Ultra-Cool and Extra-Vigorous:\\[1mm] Rotation and Activity in M and L dwarfs}

\author{A. Reiners\inst{1}\thanks{Emmy Noether Fellow}}

\titlerunning{Rotation and Activity in M and L dwarfs}
\authorrunning{A. Reiners}
\institute{Georg-August-Universit\"at, Institut f\"ur Astrophysik, Friedrich-Hund-Platz 1, D-37077 G\"ottingen}

\received{August 30, 2007}
\accepted{October 26, 2007}


\abstract{%
  The study of rotation and activity in low-mass stars or brown dwarfs
  of spectral classes M and L has seen enormous progress during the
  last years. I summarize the results from different works that
  measured activity, rotation, and sometimes magnetic fields. The
  generation of magnetic activity seems to be unchanged at the
  threshold to completely convective stars, i.e. no change in the
  efficiency of the magnetic dynamos is observed. On the other hand, a
  sudden change in the strength of rotational braking appears at the
  threshold mass to full convection, and strong evidence exists for
  rotational braking weakening with lower mass. A probable explanation
  is that the field topology changes from dipolar to small scale
  structure as the objects become fully convective.  }

\maketitle

\section{Introduction}

Rotation and activity are intimately connected in sun-like stars.
Rotation is believed to generate a magnetic field through a dynamo
mechanism that scales with rotation. The stellar wind couples to the
rotating magnetic field lines carrying away angular momentum so that
the star is being braked. Hence stars of spectral type F--K rotate
more rapidly and are more active when they are young, but they
decelerate and become less active as they age; this is the so-called
rotation-activity connection (Noyes et al., 1984; Pizzolato et al.,
2003).

The scaling of activity with rotation depends on the type of dynamo
inside the star. The fact that the rotation-activity connection is
similar in virtually all stars that harbor convective envelopes and
at all ages (Pizzolato et al., 2003; Reiners, 2007) indicate that the
dynamo mechanism is comparable in all these stars. Around spectral
type M3.5, the internal structure of the stars changes; stars later
than around M3.5 are believed to be fully convective. They cannot
harbor an interface dynamo working at the tachocline, which is
believed to be the most important dynamo mechanism in the hotter
sun-like stars. If the interface dynamo was the only important
mechanism driving a magnetic dynamo, one could expect a sharp break in
magnetic field generation around spectral type M3.5. Such a break
would imply a sudden change in observable stellar activity and in the
braking of stellar rotation.

A break in stellar activity is not observed in X-rays or H$\alpha$ in
the surveys that crossed the M3.5 border (e.g., Delfosse et al., 1998;
Mohanty \& Basri, 2003; West et al., 2004).  Activity rather stays at
comparable levels if normalized to bolometric luminosity, and quite
surprisingly the fraction of active stars even raises up to $\sim
80\,\%$ in late M-type objects before it goes down again. This clearly
shows that the interface dynamo is not the only dynamo operating in
stellar interiors and that fully convective stars can have quite
efficient dynamos as well.

On the other hand, Delfosse et al., 1998, presented a plot that shows
a different behavior in rotational braking in stars later than
spectral type M3.5. In their Fig.\,3, they show that all M dwarfs
earlier than M3.5 are slow rotators ($v\,\sin{i} \la 3$\,km\,s$^{-1}$)
regardless of what disk population they belong to (young or old). M
stars later than M3.5, however, show substantial rotation velocities
of up to $v\,\sin{i} = 50$\,km\,s$^{-1}$ in the young disk population,
and in the old disk population some late M dwarfs still have
velocities around $v\,\sin{i} = 10$\,km\,s$^{-1}$. This can be
interpreted as an indication for a sudden change in the timescales of
rotational braking at the mass where stars become fully convective.
Delfosse et al.  conclude that spin-down timescales are on the order
of a few Gyrs at spectral type M3--M4, and of the order of 10\,Gyr at
spectral type M6.

The investigation of rotation and activity in ultra-cool stars (M7 and
later) was put on firm ground by Mohanty \& Basri, 2003. From
high-resolution spectra they determined projected rotation velocities,
and from H$\alpha$ emission they derived the level of activity.
Reiners \& Basri, 2007, added more M stars to this sample. In
addition, they measured magnetic fields in low-mass M dwarfs and
showed that in M dwarfs the level of activity is still coupled to
magnetic flux -- high magnetic flux levels lead to strong H$\alpha$
emission and stars without H$\alpha$ emission show no magnetic fields.

\section{New Data}

In this paper, I summarize the measurements of rotation and activity
that are available from Delfosse et al., 1998; Mohanty \& Basri, 2003;
Reiners \& Basri, 2006. Additionally, I show first results from a new
survey in L dwarfs (Reiners \& Basri, in preparation), in which we
measured rotation and activity in almost 50 objects of spectral type
L. Our data come from the HIRES spectrograph at Keck observatory and
from UVES at ESO/VLT.

\section{Activity}

\begin{figure}
  \centering 
  \resizebox{\hsize}{!}{\includegraphics{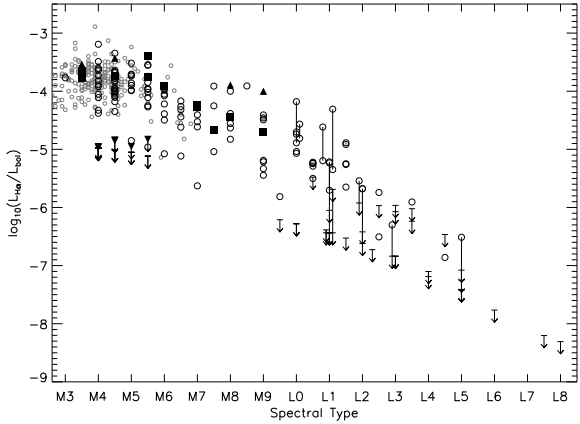}}
  \caption{\label{fig:activity}Normalized H$\alpha$ activity
    $\log{L_\mathrm{H\alpha}/L_\mathrm{bol}}$ vs. spectral type.
    Filled symbols mark stars with measured magnetic flux; upward
    triangles are stars with high magnetic flux ($Bf > 3$\,kG),
    squares denote stars with intermediate flux level ($3> Bf >
    1$\,kG), and downward triangles mean no or little magnetic flux
    $Bf < 1$\,kG. Open circles are from high resolution spectra from
    different sources (see text). Small grey circles are taken from
    Hawley et al., 1996.}
\end{figure}

Activity in low mass stars is usually measured in terms of normalized
H$\alpha$ luminosity, $\log{L_\mathrm{H\alpha}/L_\mathrm{bol}}$. It
can be calculated from the equivalent width of the H$\alpha$ line
using flux calibration from spectral models (e.g., Mohanty \& Basri,
2003). Fig.\,1 shows the values of
$\log{L_\mathrm{H\alpha}/L_\mathrm{bol}}$ for the results of Delfosse
et al., 1998; Mohanty \& Basri, 2003; Reiners \& Basri, 2006; and the
new L dwarf data. The stars for which magnetic flux measurements are
available in Reiners \& Basri, 2006, are plotted with filled symbols
distinguishing between three groups of strong, intermediate and little
magnetic flux (see caption of Fig.\,1).  All other detections of
H$\alpha$ emission are shown by open circles, non-detections by
downward arrows. I also plot the results from Hawley et al., 1996, in
small grey circles.  They come from less sensitive spectroscopy so
their detection limit is higher than limits from high resolution
spectroscopy.

The measurements from Hawley et al., 1996, are part of a large survey
among nearby M stars that extends to early M dwarfs. Normalized
activity scatters around $\log{L_\mathrm{H\alpha}/L_\mathrm{bol}} =
-4$ and shows no break at the threshold to complete convection
($\sim$M3.5). Results from the other surveys show that this activity
level is maintained up to at least M7 before the general level of
activity gradually decreases. This decrease is believed to be due to
the growing neutrality of the atmosphere (Mohanty et al., 2002) and
does not necessarily imply a less efficient dynamo mechanism.

Activity is observed in objects as cool as spectral type L5
($T_\mathrm{eff} \approx 1700$\,K). Especially among the L dwarfs, a
number of objects was observed more than once, and the data points
that belong to the same objects are connected with solid lines. Some L
dwarfs show large variability in H$\alpha$, i.e. they frequently show
flaring activity. In the few objects that are as late as L7--L8
($T_\mathrm{eff} \approx 1500$\,K), no H$\alpha$ activity was
detected. Whether this means that objects as cool as this are
inactive, or that their activity level is just below the detection
limit is an open question (see also Burgasser et al., 2002).

\section{Rotation}

\begin{figure}
  \centering 
  \resizebox{\hsize}{!}{\includegraphics{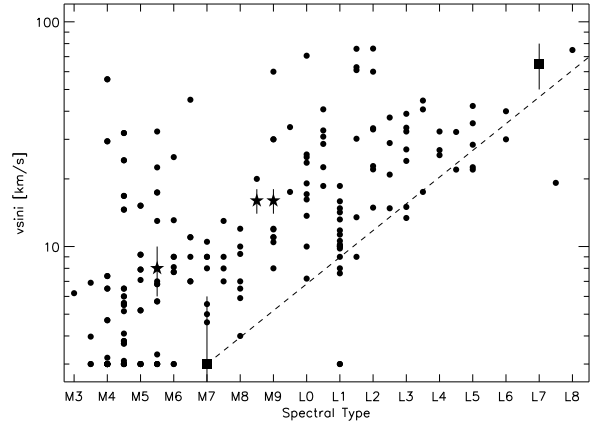}}
  \caption{\label{fig:rotation}Projected rotational velocities
    $v\,\sin{i}$ as a function of spectral type. Results from field
    star investigations (see text) are plotted as circles. Stars mark
    the three components of LHS\,1070, which are probably of same age.
    The filled squares show two sub-dwarfs that are probably as old as
    the galaxy itself. The dashed line shows the approximate lower
    envelope of rotation velocities.  }
\end{figure}

Measuring rotation velocities requires spectra of higher quality than
the detection of the H$\alpha$ emission line. A comparison to
artificially broadened synthetic spectra becomes increasingly
difficult with later spectral type, because a) the synthetic spectra
systematically provide a worse match to the spectral details as more
and more molecular features shape the spectrum (and dust begins to
form in the atmosphere), and b) the typically broad molecular features
are less affected by rotation broadening than sharp atomic absorption
lines so that the accuracy of $v\,\sin{i}$ measurements in M and L
dwarfs is usually lower than in sun-like stars.

In Fig.\,2, I show as filled circles the results of $v\,\sin{i}$
measurements in M- and L-dwarfs from Delfosse et al. (1998); Mohanty
\& Basri (2003); Reiners \& Basri (2007); and from new L-dwarf
measurements. In contrast to the early M dwarfs of spectral type
earlier than M3 (see Delfosse et al., 1998), many objects exhibit
significant rotation rates.  Mohanty \& Basri (2003) already showed
that L dwarfs are probably rapidly rotating in general, and that there
might be a lack of slowly rotating L dwarfs. With the new
observations, it has now become obvious that slow rotators are rare
among the visible L dwarfs. The dashed line in Fig.\,2 marks the lower
envelope of rotational velocities as a function of spectral type that
seems to limit the lower end of rotation rates in ultra-cool dwarfs.
The few exceptions may well be due to projection effects, i.e.  the
two ``slow'' rotators at spectral types L1 and L7.5 may be observed
pole-on.

Many of the ultra-cool dwarfs are probably brown dwarfs that do not
establish a fixed temperature, they rather keep cooling down the
spectral sequence. If they are being rotationally braked, they do not
follow vertical tracks in $v\,\sin{i}$ at constant spectral type in
Fig.\,2, but they rather evolve for example from rapidly rotating
mid-M dwarfs to slowly rotating late L dwarf as they follow the
evolutionary tracks. 

A possible scenario could be that the rotationally braked slow
rotators are old objects that are much fainter than their rapidly
rotating predecessors. The slowly rotating L dwarfs could simply be
too faint for the samples, and the dashed lower envelope could be due
to an observational bias. A few arguments though contradict the
picture of an observational bias. First, in this scenario rotational
braking of ultra-cool objects must follow quite accurately the cooling
of the objects (independent of their mass), and the distribution of
ultra-cool objects must be very homogeneous. Second, there are two
observed systems indicating that the spread of rotational velocity
with spectral type is indeed mass (or temperature) dependent, not time
dependent: 1) Reiners et al. (2007) have recently taken spectra of the
triple system LHS\,1070.  The system consists of three M dwarfs, one
mid-M dwarf of spectral type M5.5 and two late-M dwarfs.  Although the
three objects are probably coeval, the two late M objects are rotating
twice as fast as the more massive mid-M object. This indicates that at
the same age the two cooler objects went through the same rotational
history while the third, more massive component suffered stronger
rotational braking.  2) Two ultra-cool sub-dwarfs were investigated by
Reiners \& Basri (2006).  Late-type sub-dwarfs are probably metal poor
relics of the early galaxy and have had billions of years to slow down
their rotation rate. The two objects are indicated with filled squares
in Fig.\,2.  While the sdM7 dwarf does not exhibit any signs of
rotation, the sdL7 object still rotates at very high velocity,
$v\,\sin{i}\approx 65$\,km\,s$^{-1}$. This is a very strong argument
that we can indeed see the oldest L dwarfs even at late spectral
class, but that they are still rotating at very high pace.

\section{Summary and Conclusions}

High resolution spectroscopy in M and L dwarfs becomes a technique
that allows to investigate the physics and the evolution of low mass
objects in great detail. No change in H$\alpha$ activity is observed
at the threshold to complete convection. Activity can be followed to
objects as late as mid-M and it might continue to even lower masses
but below the current observational threshold. The decline in
normalized activity among the ultra-cool dwarfs could be explained by
the enhanced electrical resistivity at the low temperatures.
Currently, no indications for a mass or temperature dependence of
stellar or substellar dynamos can be concluded from the activity
measurements.

A sudden change in the behavior of rational braking at spectral class
M3.5 was already found by Delfosse et al., 1998. In the young disk,
stars earlier than M3.5 rotate slowly while later stars still show
significant rotation. The lack of slowly rotating ultra-cool dwarfs
and the rise of the minimum rotation rate with spectral class could be
explained by rotational braking that is weaker with lower mass or
lower temperature. The reason for a weaker braking at later spectral
type could be a different magnetic topology.  The geometry of the
magnetic field is essential for the strength of magnetic braking as
described in Krishnamurti et al.  (1997) and Sills et al.  (2000).
These authors find that the description of magnetic braking
($\omega_\mathrm{crit}$) needs to be different in mid-M stars and in
earlier objects, which they connect to different convective turnover
times.

Although the details of magnetic braking in ultra-cool dwarfs are not
quite understood, a substantial amount of evidence exists that
rotational braking is weaker with lower mass or lower temperature. The
limiting factor of rotational braking may be the topology of the
magnetic fields which in fully convective stars might be generated on
smaller scales so that the topology is different from a dipolar
configuration leading to the weaker rotational braking.

\acknowledgements A.R. received financial support from the DFG as an
Emmy Noether Fellow (RE 1664/4-1).

\end{document}